# A Case of Quantum Scattering

Eduardo V. Flores[1]
Department of Physics & Astronomy
Rowan University
Glassboro, NJ 08028

We analyze the results of an experimental setup that consist of two statistically independent laser beams that cross, interfere and end at detectors. At the beam intersection we place a thin wire at the center of a dark interference fringe and analyze the complementarity inequality. We find that the complementarity inequality is fulfilled provided we include a scattering interaction. We find that this interaction is implicit in the formalism of quantum theory; however, this interaction is active only when the conservation laws are satisfied.

**Introduction**

Scientist have identified gravitational, strong, weak and electromagnetic interactions. While certain aspects of these interactions are understood, there are other aspects that need more study. For instance, there is a force-like interaction that keeps identical fermions from occupying the same state. This property of fermions can result in physical effects such Fermi pressure that can be strong enough to keep neutron stars and white dwarfs from collapsing [1]. Entangled photons with net angular momentum can interact at large distances in such a way that angular momentum is conserved [2]. These interactions can be strong and sometimes even instantaneous. A related interaction is present in the formation of bright and dark interference fringes by accumulation and exclusion of particles according to corresponding constructive and destructive wave interference. The net effect of this interaction is as if a force field had acted on the particles. In this paper we present an example of scattering related to the formation of bright and dark interference fringes.

**Particle wave duality**

Particle aspect can be associated with the which-way information parameter, $K$. Wave aspect can be associated with the visibility parameter, $V$. Here, we define $K$ and $V$ in such a way that they could be applied to the experimental setup in Ref [3]. Following the work of Wheeler, we associate which-way information with the degree to which we could identify the origin of a detected particle by applying momentum conservation [4]. In the setup we analyze, a particle that could come from one of two sources, if we could determine which source generates the particle then we have maximum which-way information, $K = 1$. If we could not identify the source, we have zero which-way information, $K = 0$. We could also have partial which-way information, $1 > K > 0$. Visibility, $V$, measures the contrast between bright and dark fringes in an interference pattern formed by accumulation of particles. If we could determine that a particle avoids a given region and is attracted to an adjacent region then its visibility is maximum, $V = 1$, since it contributes to the formation of adjacent dark and bright fringes. On the other hand, if we determine that the particle is equally likely to reach any place within a region then its visibility is zero, $V = 0$, since it reflects a uniform distribution. We also could have partial visibility, $1 > V > 0$. The complementarity inequality [5,6] is given by

$$K^2 + V^2 \leq 1. \tag{1}$$

---

We note that it is not always possible to directly measure which-way information and visibility. When we cannot directly measure either of these parameters we use indirect ways such as applying the conservation laws [4].

We analyze an experiment [3] that consists of two statistically independent [7] laser beams in phase that cross at small angle, separate and end at detectors as in Fig. 1. The beams have a Gaussian profile. The experiment can be run at a photon count so low that at a given time there is high probability that only one photon is present in the entire setup [3]. In this work it is enough to use the plane wave approximation for the photons in the beams; thus, photons in beam 1 have momentum $\vec{p}_1$ and photons in beam 2 have momentum $\vec{p}_2$. Therefore, at the beam intersection there is a two-state system with momentum components $\vec{p}_1$ and $\vec{p}_2$ and their corresponding regions of constructive and destructive wave interference. The situation at the beam intersection may be summarized by the expression

(2)
$$H_0 = \delta_{\vec{p}_{\text{in}},\vec{p}_{\text{out}}}[a_1^\dagger a_1 + a_2^\dagger a_2].$$

Equation 2 expresses momentum conservation; thus, if photon $|1\rangle_1$ comes to the intersection with momentum $\vec{p}_1$, then, it leaves the intersection with the same momentum. The delta function emphasizes momentum conservation. Therefore, when detector 1 clicks we know that the photon originated at source 1 as seen in Fig. 1; thus, which-way information is maximum, $K = 1$. We note that it takes external momentum to produce deflections in the original Gaussian momentum distribution to form a Gaussian momentum distribution with interference fringes. Therefore, in this case, the original Gaussian particle distribution across the beam, which has zero visibility, $V = 0$, must be maintained from source to end detector. The complementarity inequality in Eq. 1 is not violated. We note that, in this case, momentum conservation is used to estimate both, $K$ and $V$.

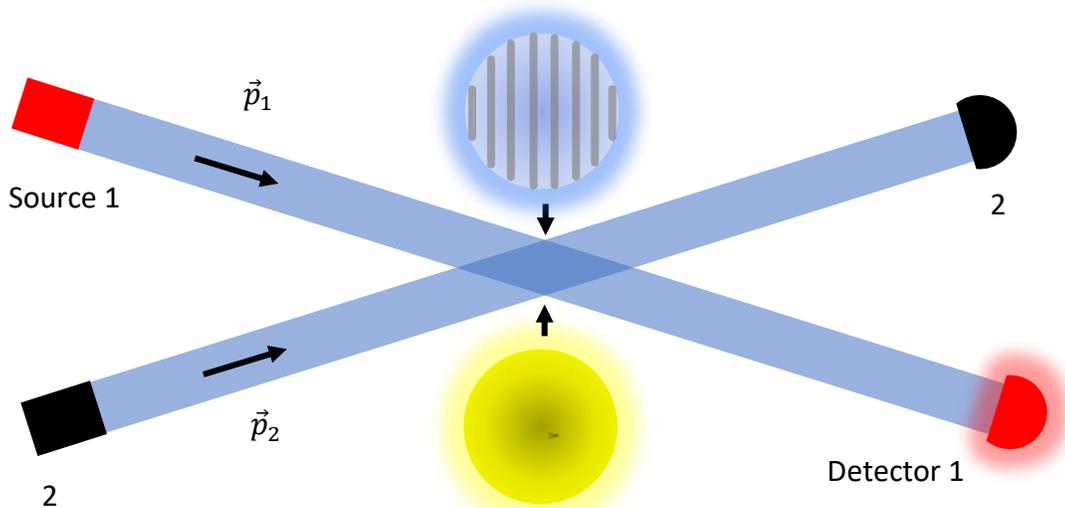

**Fig. 1 Two laser beams intersect and end at detectors.** The probability that there is only one photon in the entire setup is high. When detector 1 clicks then the origin of the detected photon is attributed to source 1; thus, we have maximum which-way information, $K = 1$. At the beam intersection there is constructive and destructive wave interfere as the pattern in blue and grey shows. However, due to momentum conservation, Gaussian momentum distribution, in yellow, must be maintained from source to end detector. Thus, the visibility attributed to these photons is zero, $V = 0$. The complementarity inequality is not violated.

When an opaque screen is placed at the beam intersection then we see an interference pattern with maximum visibility, $V = 1$. Photons end at the screen forming highly localized bright fringes in such a way that their uncertainty in momentum is greater than the difference, $\vec{p}_2 - \vec{p}_1$, which means that their origin cannot be identified [3]. Thus, for these photons, the which-way information is zero, $K = 0$. There is no violation of the inequality in Eq. 1. We note that the interference pattern on the screen shows evidence of a strong force-like field that deflects photons from a Gaussian distribution to a Gaussian distribution with bright and dark fringes. In this case, the opaque screen itself must be the source of external momentum necessary for photon deflection.

**A paradox**

In the experiment we analyze, they scan 17 $\mu$m thick wire across the beam intersection and see evidence of constructive and destructive interference depending on wire location [3]. A calculation of wire diffraction using Fraunhofer diffraction shows excellent agreement with experimental results [3, 8]. There is a QED calculation of wire diffraction for electrons [8]; here, the electron interacts with a wire-like potential barrier. The diffraction pattern from QED appears even closer to experimental evidence than Fraunhofer diffraction pattern; however, discrepancies among the two methods of calculation are small [8]. Experimental results show that when one of the beams is blocked there is no interference and decrease in photon count due to wire absorption and diffraction is as large as 8 % [3.v2]. When the two beams are on and the wire is at the center of a bright fringe decrease in photon count is as large as 12% [3]. On the other hand, when the wire is at the center of a dark fringe, wire diffraction and wire absorption are insignificant [3] and decrease in photon count at end detectors is hardly noticeable.

We analyze the case when the wire is at the center of a dark fringe. We assume that photon-wire interaction depends on distance between photon and wire. Since the wire directly affects up to 12% of the photons in the beam then about 88% of photons must pass far enough from the wire to avoid significant interaction; thus, these photons cross the beam intersection relatively free. Free photons maintain high which-way information, $K = 1$, and due to their uniform distribution across the beam their visibility is zero, $V = 0$. These photons fulfill the inequality in Eq. 1.

The remaining 12% of photons approach the wire, sitting at the center of a dark fringe, but must be redirected towards adjacent bright fringes; thus, they obtain a high level of visibility, $V \approx 1$. Calculation shows that when the wire is at the center of a dark fringe, wire diffraction is insignificant; thus, a photon that comes from source 1 is not deflected by diffraction to detector 2, instead, maintains a high level of which-way information, $K \approx 1$. Therefore, it appears that 12% of photons in the beam have high visibility and high which-way information violating the inequality in Eq. 1. The situation seems worse when a thin wire is placed at the center of every dark fringe across the beam intersection. This case has been experimentally carried out and they find that the decrease in photon count is still small, of the order of 1%; however, some of this 1% can be accounted by wire misalignment and consequent increases in wire absorption and wire diffraction [9].

**Paradox resolved**

What is needed to preserve the inequality in Eq. 1 is that out of the 12% of photons that come close to the wire, half be randomly deflected to detector 1 and half to detector 2. If this were the case then photon count would remain unchanged and the which-way information would be zero. We note that with the wire at the center of a dark fringe classical electromagnetic diffraction is not strong enough and precise enough to supply the momentum needed for photon deflection. For instance, the momentum needed to change a photon from beam 1 to beam 2 is exactly $\Delta\vec{p} = \vec{p}_2 - \vec{p}_1$.

The interaction of a photon with the wire that results in the photon either maintaining course or changing beam is given by

$$H_I = \frac{1}{\sqrt{2}}\{\delta_{\vec{p}_{in},\vec{p}_{out}}[a_1^\dagger a_1 + a_2^\dagger a_2] + \delta_{\vec{p}_{in},\vec{p}_{out}-\Delta\vec{p}}[a_2^\dagger a_1 + a_1^\dagger a_2]\}. \tag{3}$$

Here, the wire must provide the momentum necessary for deflection, $\Delta\vec{p}$. The delta functions enforce momentum conservation. For instance, if photon $|1\rangle_1$ comes to the wire with momentum $\vec{p}_1$, as seen in Fig. 2, then, it leaves in state

$$|1\rangle = \frac{1}{\sqrt{2}}\{\delta_{\vec{p}_1,\vec{p}_{out}}|1\rangle_1 + \delta_{\vec{p}_1,\vec{p}_{out}-(\vec{p}_2-\vec{p}_1)}|1\rangle_2\}. \tag{4}$$

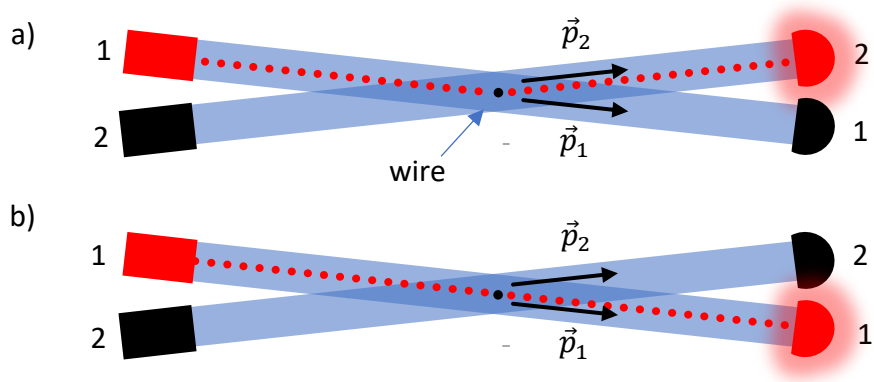

**Fig. 2 Photon scattering from a wire located at the center of a dark fringe.** We imagine a photon from source 1 heading towards the wire but somehow avoiding the wire. This photon is collaborating with the formation of a dark fringe; thus, it has high level of visibility, $V = 1$. After interacting with the wire, the photon has two possible outcomes: state $|1\rangle_1$ with momentum $\vec{p}_1$ and state $|1\rangle_2$ with momentum $\vec{p}_2$. a) If detector 2 clicks then the photon had momentum $\vec{p}_2$. We may trace the path of the photon from detector to wire. Provided that we could read the momentum $\Delta\vec{p}$ stored in the wire due to interaction with the photon, we could determine the path of the photon all the way to source 1 b) If detector 1 clicks the photon had momentum $\vec{p}_1$. If we find no excess momentum stored in the wire we could determine that the photon came from source 1. Thus, provided we could read the momentum stored in the wire, the which-way information would be maximum, $K = 1$, and the complementarity inequality would be violated, $K^2 + V^2 = 2$.

State $|1\rangle$ in Eq. 4 is not a superposition since we could in principle set the wire free to find out the momentum stored in the wire due to its interaction with the photon. Then, once a detector clicks we would be able to use momentum conservation to find the path of the photon from detector back to source, which implies full which-way information, $K = 1$. However, there are limitations imposed by the uncertainty principle. Momentum conservation shows that the momentum supplied by the wire when it deflects a photon by a small angle, $\alpha$, is $\Delta p_y \approx p\alpha$, where $p = h/\lambda$ is the photon momentum and $\alpha$ is the angle at which the beams cross. We may set the wire free and measure its momentum $\Delta p_y$. According to the uncertainty principle the uncertainty in the wire position would be $\Delta y \geq \frac{h}{\Delta p_y} = \frac{\lambda}{\alpha}$. A relatively simple calculation shows that for two monochromatic light beams with a small angle $\alpha$ between them the distance between two interference fringes at the beam intersection is given by $l = \frac{\lambda}{\alpha}$. Thus, the uncertainty in wire position, $\Delta y$, is bigger or equal to the distance between fringes, $\Delta y \geq l$. Therefore, we could not warranty that the wire is at the center of a dark fringe and the visibility would be compromised in such a way that our analysis would break down.

Since we cannot set the wire free to measure its momentum changes we firmly attach the wire to the whole setup. Now, the net momentum transferred to the photon by the whole setup is zero for both cases. When the photon gets momentum from the wire to change direction, the wire momentum is equal and opposite to the sum of momenta due to recoil of the source and detector. When the photon maintains its direction from source to detector the recoil of the source is equal and opposite to the recoil of the detector. Since no momentum is stored in the setup we cannot find out which of the two possibilities the photon took; thus, state $|1\rangle$ in Eq. 4 changes to

(5)
$$|1\rangle = \frac{1}{\sqrt{2}}\{|1\rangle_1 + |1\rangle_2\}.$$

Now, the outgoing photon has equal chance of being found in beam 1 or beam 2. When a detector clicks we do not know which source originated the photon and the which-way information is zero, $K = 0$. We call this interaction a case of quantum scattering. Therefore, photons that scatter from a firmly set wire, sitting at the center of a dark fringe, gain high visibility, $V = 1$, but loose which-way information, $K = 0$, so that there is no violation of the inequality in Eq. 1.

**Concluding remarks**
We note that when the two identical laser beams cross freely there are two momentum states at the beam intersection with corresponding regions of constructive and destructive wave interference. However, due to momentum conservation photons do not switch states and seem unaffected by the presence of constructive and destructive wave interference. Thus, particles have full which-way information and zero visibility and the complementarity inequality is not violated. However, when a macroscopic object, that could provide the necessary momentum for deflection, is present then particles follow wave interference. This is the case of an opaque screen at the beam intersection whose presence results in an interference pattern on the opaque screen. Similarly, when a thin wire is scanned across the beam intersection changes in photon count at end detectors reveal regions of constructive and destructive interference at the wire location.

Experimental results show that when the wire is at the center of a dark fringe, photon count at end detectors hardly changes; thus, there is evidence of formation of a dark fringe which implies high visibility. A classic wire diffraction calculation, for this case, shows that wire diffraction and wire absorption are insignificant; thus, photons seem undisturbed which implies high which-way information. High visibility and high which-way information would result in a violation of the complementarity inequality. However, a basic feature of quantum mechanics that we call a case of quantum scattering avoids the violation.

Our case of quantum scattering is the interaction that considers momentum exchange between photons and wire. This interaction is remarkable because it only transfers the necessary momentum for the particle to end up in a prescribed momentum state. In addition, quantum scattering introduces randomness in the final states. We note that when the wire is at the center of a dark fringe there is minimal wire diffraction and absorption. On the other hand, at this location, our case of quantum scattering has a strong effect on every photon. We notice that fermi pressure, formation of bright and dark fringes, non-local correlations and our case of quantum scattering are different facets of the same interaction. This interaction is implicit in the formalism of quantum mechanics and its physical manifestations are sometimes known as quantum effects. Unfortunately, the formalism of quantum mechanics does not seem to tell us how energy, momentum and angular momentum are transferred to produce these effects.

**References**

1. Hartle, J. B., Gravity (Addison-Wesley, San Francisco, 2003)
2. A. Aspect, J. Dalibard, and G. Roger, Experimental Test of Bell's Inequalities Using Time-Varying Analyzers, Phys. Rev. Lett. **49**, 1804
3. E. V. Flores, J. Scaturro, "Light: Particle & Wave, **arXiv:1412.1077v4**
4. J. A. Wheeler, "Law Without Law," Quantum Theory and Measurement (Princeton: Princeton University Press, 1983).
5. D. Greenberger and A. Yasin, "Simultaneous wave and particle knowledge in a neutron interferometer," Phys. Lett. A **128**, 391 (1988).
6. B.-G. Englert, "Fringe visibility and which-way information: An inequality," Phys. Rev. Lett. **77** 2154 (1996).
7. R. Loudon, The Quantum Theory of Light, 3rd Ed. (Oxford Univ. Press, 2000), p. 226.
8. A. Daniels, J. Kupec, T. Baker & E. Flores, Diffraction from static potential in QED, **arXiv:1612.00441v2**
9. S. S. Afshar, E. V. Flores, K. F. McDonald, E. Knoesel, "Paradox in particle-wave duality," Found. Phys., **37** (2) 295-305 (2007).